\documentclass[12pt]{article}
\usepackage{graphicx} 
\usepackage{mathrsfs,amsmath,amssymb,physics,amsfonts,latexsym,mathtools, tensor,color,framed,ascmac,xcolor,here}
\usepackage{blkarray}
\usepackage[top=25truemm,bottom=20truemm,left=20truemm,right=20truemm]{geometry}
\usepackage{hyperref}
\usepackage{amsthm}
\usepackage{tikz}
\usetikzlibrary{arrows.meta, decorations.pathreplacing}
\usepackage{appendix}

\theoremstyle{definition}

\newtheorem*{thm*}{lemma}

\numberwithin{equation}{section}
\renewcommand\theequation{\arabic{section}.\arabic{equation}}

\title{Topology of Calorons Re-examined}
\author{Atsushi Nakamula\footnote{e-mail: nakamula@sci.kitasato-u.ac.jp} and Genki Sumiyama\footnote{e-mail: sumiyama.genki@st.kitasato-u.ac.jp
} {}\footnote{Corresponding author} \\
\\
  Department of Physics, School of Science, Kitasato University\\
  Sagamihara 252-0373, Japan
}
\date{}

\begin{document}




\maketitle

\begin{abstract}
We reconsider the detailed structure of the topological character of the instantons in pure Yang-Mills theory on $S^1\times\mathbb{R}^3$, known as calorons.
We argue that the standard formula for the second Chern number requires modification, as revealed by explicit analytic consideration.
For concreteness, we compute the second Chern number of the Harrington-Shepard caloron of the two-dimensional special unitary gauge group ($\mathrm{SU}(2)$) with unit topological charge in several gauges.
The general formula is shown to remain valid even when the gauge connection is in a singular gauge.
The gauge dependence of the magnetic charge is also discussed. 
\end{abstract}

\section{Introduction}

Topological objects in field theories \cite{Manton:2004tk,Shifman:2009zz,Weinberg:2012pjx,Shifman:2012zz} yield a multitude of curious non-perturbative phenomena.
In four-dimensional Yang-Mills gauge theory, it is believed that these objects, such as instantons, monopoles, and vortices, play a crucial role in governing color confinement \cite{Wilson:1974sk}, chiral symmetry breaking \cite{Nambu:1961fr,Nambu:1961tp}, and so on.
In recent years, intensive studies have been carried out on these topological objects in Yang-Mills theories of an $N$-dimensional special unitary group ($\mathrm{SU}(N)$) defined on circle-compactified spaces, namely, $S^1\times \mathbb{R}^3,\,T^2\times\mathbb{R}^2,\,T^3\times\mathbb{R}$ and $T^4$ in efforts to understand the descriptions of such non-perturbative phenomena in pure Yang-Mills or super-Yang-Mills theories \cite{Tanizaki:2017qhf,Tanizaki:2022ngt,Anber:2022qsz,Anber:2023sjn,Anber:2024uwl,Hayashi:2025doq}.

A distinctive feature of these topological objects on the circle compactified spaces is the possibility that they possess non-integer, or fractional, topological numbers.
For SU($N$) instantons on a hypertorus $T^4$, a fractional topological number can arise due to a twisted boundary condition, the so-called 't Hooft twist \cite{tHooft:1981nnx,tHooft:1981sps,vanBaal:1982ag,Gonzalez-Arroyo:2023kqv}. 
Although this was discovered in the 1980s, fractional instantons on $T^4$ have re-emerged as a key issue in studies on generalised global symmetries \cite{Gaiotto:2014kfa,Gaiotto:2017yup}.
Furthermore, numerous studies are actively being pursued on monopoles on $S^1\times\mathbb{R}^3$ and center vortices on $T^2\times\mathbb{R}^2$ to explore color confinement at the semiclassical level, e.g., \cite{Guvendik:2024umd,Hayashi:2025mgk,Kondo:2025dtx}.  
Note that, in this context, the compactified circle and torus directions are non-thermal, i.e., they are compactified spatial directions, in contrast to the thermal compactification of the Euclidean time direction that characterises finite-temperature field theory. 
Although these topological objects are adiabatically connected to the fractional instantons on $T^4$ through appropriate limits of the circle lengths, the mutual relationships among these objects on distinct compactified spaces remain unclear.

The central concern of the present paper is the instantons on $S^1\times\mathbb{R}^3$, known as calorons within the framework of thermal field theory.
Focusing mainly on the topological characterization of calorons, we re-examine the formula of the second Chern number in a gauge-independent manner, and find a modification to the results of a previous study on the topological invariant \cite{Nye:2000eg,Nye:2001hf}, whose primary interest was the index theorem in the background of calorons.
As is well known, the Bogomolnyi-Prasad-Sommerfield (BPS) monopoles \cite{Prasad:1975kr} are interpreted as calorons with translational invariance along the $S^1$ direction.
Moreover, since calorons reduce to instantons in $\mathbb{R}^4$ in the infinite-period limit of $S^1$, the topological character of calorons has a rich structure involving both magnetic and instanton charges.  
Although calorons carry non-integer topological numbers in general, we focus here on the integer case, namely calorons of Harrington-Shepard type \cite{Harrington:1978ve} as a first step toward establishing the general formula for the topological characterisation of calorons.
The non-integer, or non-trivial holonomy, cases will be addressed in a subsequent paper.
We emphasise that we do not distinguish here between a spatial compactified circle and an imaginary-time one -- that is, between non-thermal and thermal compactifications -- and investigate only the topological character of instantons on $S^1\times\mathbb{R}^3$ at the classical level.

In the next section, we review the previous study on the topological character of calorons and argue that a refinement is needed.
This claim is verified in Section III, where an analytic calculation is performed for a caloron gauge field in several distinct gauges as a concrete illustration.
We make use of elements of Sato's hyperfunction theory, which is reviewed in the Appendix.
The final section is devoted to a summary.

\section{Topological Character of Calorons}\label{Sec: Topological Character}

A pure Yang-Mills action is bounded below by the second Chern number when the field strength tensor, or equivalently the curvature two-form, satisfies the selfdual or anti-selfdual equations.
The (anti-)selfdual equations relating the field strength tensor to its Hodge dual are
\begin{align}\label{ASD}
    F_{\mu\nu}=\pm\tilde{F}_{\mu\nu}=\pm\frac{1}{2}\varepsilon_{\mu\nu\lambda\rho}F_{\lambda\rho},
\end{align}
where the Greek indices run from $0$ to $3$, $\varepsilon_{\mu\nu\lambda\rho}$ is the completely antisymmetric tensor with $\varepsilon_{1230}=-\varepsilon_{0123}=1$, and the field strength tensor is defined from a gauge potential $A_\mu$ as,
\begin{align}\label{field strength}
    F_{\mu\nu}=\partial_\mu A_\nu-\partial_\nu A_\mu+[A_\mu,A_\nu].
\end{align}
Throughout this paper, we consider the anti-selfdual (the minus sign) case, and raise and lower indices freely, since our analysis is conducted entirely in Euclidean flat space.

The second Chern number for Yang-Mills gauge field configurations on $S^1\times\mathbb{R}^3$, 
\begin{align}\label{2nd Chern number}
     Q=-\frac{1}{8\pi^2}\int_{S^1\times\mathbb{R}^3}\tr F\wedge F,
 \end{align}
provides the topological classification of the field configuration, where $F$ is the Lie algebra valued curvature two-form defined from the connection one-form $A$ by $F=dA+A\wedge A$, an equivalent definition to \eqref{field strength}, and the trace is taken over the Lie algebra $\mathfrak{g}$ of the gauge group $G$.
We use anti-Hermitian basis of $\mathfrak{g}$ throughout.
Let the circumference of $S^1$ be $\beta=2\pi/\mu_0$, where $\mu_0$ is the period of the dual space introduced for later use.
In the context of finite-temperature field theories, $\beta$ represents the inverse temperature.
Hereafter, we adopt Cartesian coordinates on Euclidean space in the notation $S^1\times\mathbb{R}^3\ni x=(x_0,x_1,x_2,x_3)=:(x_0,\vec{x})$.

The explicit integration formula of \eqref{2nd Chern number} has been established by Nye and Singer in \cite{Nye:2000eg,Nye:2001hf}, which reads for gauge group $G=\mathrm{SU}(N)$,
\begin{align}\label{Nye-Singer formula}
    Q&=-\frac{1}{24\pi^2}\int_{\mathbb{R}^3\simeq S^3}\tr(dcc^{-1})^3-\frac{1}{8\pi^2}\int_{S^1\times S^2_\infty}\tr 2F\wedge A_0\nonumber\\
    &=n+\frac{2}{\mu_0}\sum_{j=1}^{N-1}\mu_jk_j,\quad n,k_j\in\mathbb{Z},
\end{align}
where $(dcc^{-1})^3:=dcc^{-1}\wedge dcc^{-1}\wedge dcc^{-1}$ and $c=c(\vec{x})$ is a clutching function to be defined shortly,  and $k_j$'s and $\mu_j$'s are a set of magnetic charges and of holonomy parameters associated with each caloron configuration, respectively.
In this paper, we reconsider the formula \eqref{Nye-Singer formula} and show that it requires modification to apply to arbitrary caloron gauge configurations in various gauges, focusing on the case without holonomy parameters as a first step.
The need for this modification is justified by considering explicit examples of $\mathrm{SU}(2)$ calorons in the next section.
Note that the sign of the first term of \eqref{Nye-Singer formula} differs from that in \cite{Nye:2000eg,Nye:2001hf} due to a difference in the convention for the completely antisymmetric tensor, $\varepsilon_{0123}=1$ in those references.

We now follow the derivation of \eqref{Nye-Singer formula} in some detail along the lines of \cite{Nye:2000eg}.
First, we introduce the clutching function of a caloron configuration.
When a caloron gauge connection transferred along the $S^1$ direction once, say, from $x_0=-\beta/2$ to $\beta/2$ at a fixed $\vec{x}\in\mathbb{R}^3$, the relative relation of the connection is  
\begin{align}\label{clutching function i}
    A_i(\beta/2,\vec{x})&=c(\vec{x})A_i(-\beta/2,\vec{x})c^{-1}(\vec{x})+c(\vec{x})\partial_i c^{-1}(\vec{x}),\\
    A_0(\beta/2,\vec{x})&=c(\vec{x})A_0(-\beta/2,\vec{x})c^{-1}(\vec{x}),\label{clutching function 0}
\end{align}
with a gauge group element $c(\vec{x})\in G$, so-called the clutching function,
meaning that the caloron gauge connections are periodic with respect to $S^1$ up to a gauge transformation.
The clutching function at the spatial infinity $|\vec{x}|\to\infty$ is reasonably assumed to be a constant element of $G$.
Note that the form of the clutching functions strictly depends on the choice of gauges.

We next show the integral of the second Chern number \eqref{2nd Chern number} is bisected due to the direct product structure of the base space $S^1\times\mathbb{R}^3$.
A gauge field is well-defined on a simply-connected coordinate patch: otherwise, the single-valuedness of the field fails due to, e.g., the quasi-periodicity \eqref{clutching function i} and \eqref{clutching function 0}.
Since the circle $S^1$ is covered by at least two coordinate patches, we define the following two patches
\begin{align}\label{patches of S^1}
    \mbox{Patch 1}:x_0\in [-\frac{\beta}{2}+\varepsilon,\frac{\beta}{2}-\varepsilon], \quad
    \mbox{Patch 2}:x_0\in [\frac{\beta}{2}-\delta,-\frac{\beta}{2}+\delta],
\end{align}
for very small $\varepsilon$ and $\delta$ with $0<\varepsilon<\delta$, i.e., the patch 1 covers almost the whole $S^1$.
Taking the limit $\varepsilon,\delta\to0$, we identify the $S^1$ with $[-\beta/2,\beta/2]$ hereafter.
We thus evaluate the second Chern number \eqref{2nd Chern number} on $[-\beta/2,\beta/2]\times\mathbb{R}^3$ by using the Stokes theorem,
\begin{align}\label{2nd Chern Stokes}
   Q&=-\frac{1}{8\pi^2}\int_{[-\beta/2,\beta/2]\times \mathbb{R}^3}\tr F\wedge F
    =-\frac{1}{8\pi^2}\int_{[-\beta/2,\beta/2]\times\mathbb{R}^3}d\tr\left(dA\wedge A+\frac{2}{3}A\wedge A\wedge A\right)\nonumber\\
    &=-\frac{1}{8\pi^2}\int_{\partial([-\beta/2,\beta/2]\times\mathbb{R}^3)}\tr\left(dA\wedge A+\frac{2}{3}A\wedge A\wedge A\right)=C+K,
\end{align}
where
\begin{align}\label{Clut integral}
    C:=-\frac{1}{8\pi^2}\int_{\mathbb{R}^3}\tr\left(dA\wedge A+\frac{2}{3}A\wedge A\wedge A\right)\Bigg|^{\beta/2}_{x_0=-\beta/2}\ ,
\end{align}
and
\begin{align}\label{K integral}
    K:=-\frac{1}{8\pi^2}\int_{[-\beta/2,\beta/2]\times(\partial\mathbb{R}^3)}\tr\left(dA\wedge A+\frac{2}{3}A\wedge A\wedge A\right),
\end{align}
Notice that the second Chern number $Q$ itself is, of course, gauge invariant; however, neither $C$ nor $K$ is individually gauge invariant.
Therefore, the evaluation of these quantities must carefully account for their gauge dependence.
We refer to the integrals \eqref{Clut integral} and \eqref{K integral} as the C-integral and K-integral, respectively.

The C-integral \eqref{Clut integral} is evaluated to be
\begin{align}
    C&=-\frac{1}{24\pi^2}\int_{\mathbb{R}^3}\tr(dcc^{-1})^3+\frac{1}{8\pi^2}\int_{\mathbb{R}^3}d\tr A\big|_{x_0=-\beta/2}\wedge dc^{-1}c\nonumber\\
    &=-\frac{1}{24\pi^2}\int_{S^3}\tr(c^{-1}dc)^3-\frac{1}{8\pi^2}\int_{\partial\mathbb{R}^3}\tr A\big|_{x_0=-\beta/2}\wedge c^{-1}dc,\label{Clut integral result}
\end{align}
where the identification $\mathbb{R}^3\simeq S^3$ in the first integral is a consequences of $c(\infty)$ being constant.
The second integral of \eqref{Clut integral result} is ignored in \cite{Nye:2000eg}, for the same reason about the clutching function.
Although there exists no non-vanishing example of this term, we retain it here; the justification is given later with an illustration.
Thus, the contribution to the integral \eqref{Clut integral result} is almost coming from the first term, which takes values in the homotopy group $\pi_3(S^3)=\mathbb{Z}$ for $G=\mathrm{SU}(N)$.

The evaluation of the K-integral \eqref{K integral} is
\begin{align}\label{K integral result}
     K=-\frac{1}{8\pi^2}\int_{S^1\times(\partial\mathbb{R}^3)}\tr 2 F\wedge A_0+\tr \partial_0A\wedge A+ d \tr A_0\wedge A,
\end{align}
where we used the notation $S^1=[-\beta/2,\beta/2]$.
In \cite{Nye:2000eg}, the spatial boundary $\partial\mathbb{R}^3$ is taken to consist solely of the two-sphere at infinity $S^2_\infty$, because only the smooth gauge fields inside $\mathbb{R}^3$ are assumed.
However, in several gauges, the gauge connections have singularities in $\mathbb{R}^3$. 
For such cases, we need to regularise the integration \eqref{K integral} by excising small spheres around these singularities. 
We therefore have to consider these small spheres as additional boundaries of $\mathbb{R}^3$ besides $S^2_\infty$, as we will see in the next section. 
This is also a reason for not retaining the second integral in \eqref{Clut integral result}.

The first term on the right-hand side of \eqref{K integral result} gives the ``magnetic charges" of calorons.
We refer to this term as the FA-part of the K-integral, which reads
\begin{align}\label{FA-part}
   \mbox{(FA-part)}:=-\frac{1}{8\pi^2}\int_{S^1\times(\partial\mathbb{R}^3)}2\tr F\wedge A_0=-\frac{2}{\mu_0}\sum_{j=1}^{N-1}\mu_jk_j,
\end{align}
with $0\leq\mu_1\leq\cdots\leq\mu_{N-1}\leq\mu_0/2$.
In general, the magnetic charges appear when the scale parameters of calorons are large.

We refer to the second term of the right-hand side of \eqref{K integral result} as the dAA-part of the K-integral,
\begin{align}\label{dAA-part}
    \mbox{(dAA-part)}:=-\frac{1}{8\pi^2}\int_{S^1\times(\partial\mathbb{R}^3)}\tr \partial_0A\wedge A.
\end{align}
This term is also ignored in \cite{Nye:2000eg}, because the gauge connections were assumed to have no $x_0$-dependence at $S^2_\infty$.
However, there are legitimate reasons not to disregard this term. 
The first is, as we mentioned above, that the boundary of $\mathbb{R}^3$ is not only $S^2_\infty$ when the gauge connection has singularities; the gauge field would then have $x_0$-dependence near these singular points.
The second reason is that there are cases in which the integrand of \eqref{dAA-part} has $x_0$-dependence at $S^2_\infty$ even when the gauge field is smooth in $\mathbb{R}^3$.
In the next section, we will show that there are indeed cases in which this dAA-part \eqref{dAA-part} contributes a non-zero topological term.

The third term in \eqref{K integral result} is a total-derivative term with respect to $\partial \mathbb{R}^3$, which obviously vanishes by the Stokes theorem.

To summarise this section, the integral formula \eqref{Nye-Singer formula} of the second Chern number of calorons requires an additional contribution, at least from the dAA-part \eqref{dAA-part}.
In the next section, we show that this modification is indeed necessary, using a simple example of an analytic caloron gauge field.

\section{Concrete Calculation of Topological Character}

In this section, we perform the integration of the second Chern number of calorons derived in the previous section explicitly.
We restrict the gauge group $G=\mathrm{SU}(2)$ and the anti-selfdual gauge fields to be the so-called Harrington-Shepard types for simplicity and concreteness.

\subsection{Analytic form of caloron gauge fields}
\subsubsection{The 't Hooft ansatz and Lorenz --'t Hooft gauge}
The simplest kind of caloron configurations is given by the 't Hooft ansatz for multi-instantons in $\mathbb{R}^4$, 
 \begin{align}\label{'tHooft ansatz}
    A_\mu(x)=\frac{i}{2}\overline{\eta}_{\mu\nu}\partial_\nu\ln\phi(x),
 \end{align}
where $x:=(x_0,x_1,x_2,x_3)$ is four-dimensional Cartesian coordinate of $\mathbb{R}^4$ or $S^1\times\mathbb{R}^3$, and $\bar{\eta}_{\mu\nu}$ is the anti-selfdual 't Hooft symbol defined by
\begin{align}
    \overline{\eta}_{\mu\nu}&=\overline{\eta}_{\mu\nu}^a\sigma^a,\\
    \overline{\eta}_{\mu\nu}^a&=\varepsilon_{a\mu\nu 0}+\delta_{a\mu}\delta_{\nu 0}-\delta_{a\nu}\delta_{\mu 0},
\end{align}
and $\sigma^a$'s are Pauli matrices, $a$ runs from $1$ to $3$.
Substituting \eqref{'tHooft ansatz} into the anti-selfdual equations \eqref{ASD}, we find the four-dimensional Laplace equation for $\phi(x)$,
\begin{align}\label{4d Laplace}
    \partial_\mu\partial_\mu\phi(x)=0.
\end{align}
Since the 't Hooft symbol is anti-symmetric with respect to the space indices, the gauge connection obeys the Lorenz gauge condition 
\begin{align}
    \partial_\mu A_\mu=0.\label{Lorenz gauge}
\end{align}
The four-dimensional Laplace equation \eqref{4d Laplace} is solved by 
\begin{align}\label{multi-instantons}
    \phi(x)=1+\sum_j\frac{\lambda_j^2}{|x-a_j|^2},
\end{align}
in $\mathbb{R}^4$, where $|x-a_j|^2:=(x^\mu-a_j^\mu)(x^\mu-a_j^\mu)$, and $\lambda_j$'s and $a_j$'s are the scale parameters and the four-dimensional position vectors of each instanton, respectively.
Note that the gauge connection of multi-instantons \eqref{'tHooft ansatz} would be singular at each instanton position from the singularities of \eqref{multi-instantons}, in fact, this gives a multi-instanton of singular gauge.
We can arrange a one-caloron configuration by placing equi-scale  instantons on an axis, say, $x_0$, with period $\beta$ \cite{Harrington:1978ve}, 
    \begin{align}\label{one-caloron phi}
    \phi=&\;1+\sum_{n\in\mathbb{Z}}\frac{\lambda^2}{r^2+(x_0+n\beta)^2}\nonumber\\
    =&\;1+\frac{\lambda^2
                \mu_0}{2r}\frac{\sinh\mu_0r}{\cosh\mu_0r-\cos\mu_0x_0},
    \end{align}
where $r:=\sqrt{x_1^2+x_2^2+x_3^2}$ is the radial coordinate of $\mathbb{R}^3$, and we replaced $\beta=2\pi/\mu_0$ in the final description.
Note again that \eqref{one-caloron phi} is singular at the origin $x=(0,\overrightarrow{0})$, so the gauge connection is also singular there.
Similar construction for multi-calorons with multiple singularities is also possible in this setting. 
From \eqref{one-caloron phi}, we obtain the analytic representation of one-caloron gauge connection field in the Lorenz gauge \eqref{Lorenz gauge} from the 't Hooft ansatz \eqref{'tHooft ansatz},
\begin{align}\label{A_i of HS1 in LtH}
        A_j^{(\mathrm{LtH})}(x)&=\frac{-i\mu_0\lambda^2}{4N^{\mathrm{(LtH)}}r^2(\cosh\mu_0r-\cos\mu_0x_0)^2}\left(\frac{\mu_0\sinh\mu_0r\sin\mu_0x_0}{r}\sigma^j+f_\times(r,x_0)(\hat{\boldsymbol{x}}\times \boldsymbol{\sigma})^j\right),\\
        A_0^{(\mathrm{LtH})}(x)&=\frac{-i\mu_0\lambda^2}{4N^{\mathrm{(LtH)}}r^2}\left(\frac{-r\sinh^2\mu_0r}{(\cosh\mu_0r-\cos\mu_0x_0)^2}+\frac{\mu_0r\cosh\mu_0r-\sinh\mu_0r}{\cosh\mu_0r-\cos\mu_0x_0}\right)(\hat{\boldsymbol{x}}\cdot\boldsymbol{\sigma}),
        \label{A_0 of HS1 in LtH}
\end{align}
where the Latin index denotes the component of $\mathbb{R}^3$, $\hat{\boldsymbol{x}}$ is normalized position vector of $\mathbb{R}^3$, $\boldsymbol{\sigma}=(\sigma_1,\sigma_2,\sigma_3)$, and
\begin{align}\label{f_times of LtH}
    f_\times (r,x_0)&=\frac{\sinh\mu_0r}{r^2}(\cosh\mu_0r-\sinh\mu_0r)-\frac{\mu_0}{r}(1-\cosh\mu_0r\cos\mu_0x_0),\\
    N^{\mathrm{(LtH)}}(r,x_0)&=1+\frac{\mu_0\lambda^2\sinh\mu_0r}{2r(\cosh\mu_0r-\cos\mu_0x_0)}.\label{N of LtH}
\end{align}
The gauge field of \eqref{A_i of HS1 in LtH} and \eqref{A_0 of HS1 in LtH} is known as the Harrington-Shepard one-caloron (HS1), and we refer to it as HS1 in  Lorenz --'t Hooft gauge, labelled as (LtH).
Observe that the gauge connection in the Lorenz --'t Hooft gauge is exactly periodic along $S^1$, 
\begin{align}
    A_\mu^{(\mathrm{LtH})}(x_0+n\beta,\vec{x})=A_\mu^{(\mathrm{LtH})}(x_0,\vec{x}), \ n\in\mathbb{Z}.
\end{align}
Therefore, the clutching function for this gauge belongs to the center of $\mathrm{SU}(2)$, i.e., $c(\vec{x})=\pm1$.

\subsubsection{The Nahm construction and magnetic gauge}
The 't Hooft ansatz \eqref{'tHooft ansatz} clearly does not provide the most general form of caloron gauge fields.
To find general configurations in general gauges of calorons, we may apply 
the Nahm construction \cite{Nahm1984}, which is the analogue of the Atiyah-Drinfeld-Hitchin-Manin (ADHM) construction of Yang-Mills instantons \cite{Atiyah:1978ri} for the $S^1$ compactified space.
The general caloron configurations have holonomy parameters $\mu_j$'s, appearing in \eqref{Nye-Singer formula}, satisfying $0\leq \mu_1\leq\dots\leq \mu_{N-1}\leq\mu_0/2$ for $G=\mathrm{SU}(N)$; thus, for $G=\mathrm{SU}(2)$, there is only one holonomy parameter $\mu_1$.
In this paper, we focus on calorons with no holonomy parameter, i.e., we consider the case $\mu_1=0$, which includes the Harrington-Shepard calorons derived in the previous subsection.
Note that the cases of $\mu_1=0$ and $\mu_1=\mu_0/2$ are equivalent up to a class of gauge transformation called the rotation map \cite{Nye:2001hf,Cork:2017hnj}.

In the Nahm construction, the caloron gauge connections are obtained from a normalised ``Weyl spinor" $V$ defined on the dual circle of circumference $\mu_0$ parametrised by the coordinate $s$,
\begin{align}\label{Weyl spinor}
    V=\frac{1}{\sqrt{N}}\begin{pmatrix}
        U\\
        \boldsymbol{v}(s)
    \end{pmatrix},
\end{align}
where $U$ and $\boldsymbol{v}(s)$ are a quaternion and a $k$-dimensional spinor with quaternion entries, referred to as the ``boundary" and the ``bulk" components of the Weyl spinor, respectively, and $N$ is a normalization factor.
A caloron gauge connection is derived from the normalised Weyl spinor as
\begin{align}\label{A from Weyl spinor}
    A=(V,dV):=\frac{1}{N}U^\dagger dU+\frac{1}{N}\int_{\widehat{S^1}}\boldsymbol{v}^\dagger(s)d\boldsymbol{v}(s)
    -\frac{dN}{2N}
\end{align}
where $d$ is the derivative operator on the configuration space $S^1\times\mathbb{R}^3$ and the integration is performed over the dual circle denoted $\widehat{S^1}$.
We outline the defining conditions for the Weyl spinor $V$ below.

In this construction, the entire information of the field configuration is contained in the Nahm data, $T_\alpha(s)$ with $\alpha=0,1,2,3$ and $W$.
The former are $k\times k$ Hermitian-matrix valued periodic functions defined on the dual circle $\widehat{S^1}$ called the bulk data, and the latter is a $k$-dimensional row vector of quaternion entries, called the boundary data. 
If these data satisfy the following conditions, they give a caloron configuration and are called the caloron Nahm data.
Letting the interval of the dual circle $\widehat{S^1}$ be $s\in[-\mu_0/2,\mu_0/2]$, the conditions are: (i) the bulk Nahm equations,
\begin{align}\label{bulk Nahm eq}
    \frac{d}{ds}T_j(s)-i[T_0(s),T_j(s)]-\frac{i}{2}\varepsilon_{jkl}[T_k(s),T_l(s)]=0,
\end{align}
where the Latin indices run from $1$ to $3$, and $\varepsilon_{ijk}$ is the Levi-Civita symbol; and (ii) the ``reality conditions" ${}^tT_\alpha(s)=T_\alpha(-s)$.
Condition (iii) is for the boundary data,
\begin{align}\label{boundary Nahm eq}
    \frac{1}{2}\tr\sigma_j W^\dagger W=T_j\left(-\frac{\mu_0}{2}\right)-T_j\left(\frac{\mu_0}{2}\right),
\end{align}
where the trace is taken over the quaternion components.
We refer to \eqref{bulk Nahm eq} and \eqref{boundary Nahm eq}, together with the reality conditions, as the caloron Nahm equations.
Note that some modification of the Nahm equations is necessary for cases with a non-zero holonomy parameter $\mu_1$.

From the Nahm data, the Weyl spinor is determined as follows.
First, the bulk component $\boldsymbol{v}(s)$ is determined by solving the differential equation,
\begin{align}\label{bulk Weyl}
   \left( i\frac{d}{ds}+T_0+T_j(s) (i\sigma_j)+\mathbf{x}^\dagger\right)\boldsymbol{v}(s)=0,
\end{align}
where $\mathbf{x}=x_\mu e_\mu$ with quaternion basis $e_\mu=(e_0,e_1,e_2,e_3)=(1,-i\sigma_1,-i\sigma_2,-i\sigma_3)$.
The boundary component $U$ is then determined from the boundary values of the bulk components as,
\begin{align}\label{boundary Weyl}
   i W^\dagger U=\boldsymbol{v}\left(-\frac{\mu_0}{2}\right)-\boldsymbol{v}\left(\frac{\mu_0}{2}\right),
\end{align}
and finally, $U$ and $\boldsymbol{v}(s)$ are normalized by imposing the normalization condition $(V,V)=1$.
Note that the Weyl spinor has a redundancy under multiplication by an $\mathrm{SU}(2)$ element $g$ from the right, $V\to V'=Vg$, since this does not affect \eqref{bulk Weyl} and \eqref{boundary Weyl} and gives rise to a gauge transformation of the connection $A$ defined by \eqref{A from Weyl spinor}.

We now apply this formulation to construct the Harrington-Shepard caloron HS1 considered in the previous subsection.
For this configuration, the bulk Nahm data is given by one-dimensional Hermitian matrices, i.e., a constant four-dimensional vector of real entries,
\begin{align}
    T_\mu(s)=\alpha_\mu,
\end{align}
corresponding to the position of the caloron, which obviously satisfies the bulk Nahm equations \eqref{bulk Nahm eq} and the reality condition.
Without loss of generality, we choose $\alpha_\mu=0$ here.
The right-hand side of the boundary Nahm equation \eqref{boundary Nahm eq} then vanishes, so the boundary Nahm data is given by an arbitrary quaternion.
We take it to be a real quaternion $W=\sqrt{\mu_0}\lambda\in\mathbb{R}$, where $\lambda$ is identified with the scale parameter of the caloron appearing in the 't Hooft ansatz.
The inclusion of $\sqrt{\mu_0}$ is necessary in order to have a smooth limit to the instanton on $\mathbb{R}^4$.
More generally, the bulk Nahm data for the Harrington-Shepard $k$-caloron is given by $k\times k$-diagonal matrices of real entries, whose $(m,n)$-components are
\begin{align}\label{HSk bulk data}
    (T_\mu)_{mn}=\alpha_\mu^{(m)}\delta_{mn},
\end{align}
with $1\leq m,n\leq k$, and $\alpha_\mu^{(m)}$ are real four-vectors corresponding to the positions of each caloron.

Focusing on HS1, the bulk component of the Weyl equation \eqref{bulk Weyl} becomes a one-component quaternion-valued ``spinor", which is solved by
\begin{align}\label{Weyl bulk component of Mag}
    \boldsymbol{v}(s)=e^{i\boldsymbol{x}^\dagger s}=e^{ix_0s}(\cosh rs-\sinh rs \;\hat{\boldsymbol{x}}\cdot\boldsymbol{\sigma}),
\end{align}
and the boundary component is determined from the boundary Weyl equation \eqref{boundary Weyl} as,
\begin{align}\label{Weyl boundary component of Mag}
    U&=-\frac{2}{\sqrt{\mu_0}\lambda}\left(\sin\frac{\mu_0 x_0}{2}\cosh\frac{\mu_0 r}{2}+i\cos\frac{\mu_0 x_0}{2}\sinh\frac{\mu_0 r}{2}\;\hat{\boldsymbol{x}}\cdot\boldsymbol{\sigma}\right)\nonumber\\
    &=:-\frac{2}{\sqrt{\mu_0}\lambda}\mathcal{Q},
\end{align}
where the quaternion $\mathcal{Q}$ of absolute value
\begin{align}\label{abs of Q}
    |\mathcal{Q}|=\sqrt{\frac{\cosh \mu_0 r-\cos\mu_0 x_0}{2}},
\end{align}
is defined for later use.
The normalization factor, denoted $N^{\mathrm{(Mag)}}$, is determined from $(V,V)=1$ as
\begin{align}\label{normalization of Mag}
    N^{\mathrm{(Mag)}}=1+\frac{2r(\cosh \mu_0 r-\cos\mu_0 x_0)}{\mu_0\lambda^2 \sinh\mu_0 r},
\end{align}
Here, we refer to the Weyl spinor obtained as that of HS1 in the ``magnetic gauge" labelled (Mag), for the purpose of classifying gauges. 
Hence, the Weyl spinor in the magnetic gauge is
\begin{align}\label{Weyl of mag}
    V^{\mathrm{(Mag)}}=\sqrt{\frac{r}{N^{\mathrm{(Mag)}}\sinh\mu_0r}}
    \begin{pmatrix}
        -2\mathcal{Q}/\sqrt{\mu_0}\lambda\\
        \boldsymbol{v}(s)
    \end{pmatrix}.
\end{align}
Substituting \eqref{Weyl of mag} into \eqref{A from Weyl spinor}, we find the gauge connection of HS1 in a gauge different from \eqref{A_i of HS1 in LtH} and \eqref{A_0 of HS1 in LtH},
\begin{align}\label{A_i of HS1 in Magnetic}
    A_j^{(\mathrm{Mag})}(x)&=\frac{-i}{\mu_0\lambda^2N^{(\mathrm{Mag})}}\left\{-\sin\mu_0x_0 \sigma^j+\frac{(1+\cos\mu_0x_0)(\cosh\mu_0r-1)}{\sinh\mu_0r}(\hat{\boldsymbol{x}}\times \boldsymbol{\sigma})^j\right.\nonumber\\
    &\left.-\sin\mu_0x_0\left(\frac{\mu_0r}{\sinh\mu_0r}-1\right)\hat{x}^j(\hat{\boldsymbol{x}}\cdot\boldsymbol{\sigma})\right\}+\frac{-i}{N^{(\mathrm{Mag})}}\left(\frac{1}{2r}-\frac{\mu_0}{2\sinh\mu_0r}\right)(\hat{\boldsymbol{x}}\times \boldsymbol{\sigma})^j,\\
    A_0^{(\mathrm{Mag})}(x)&=\frac{-i}{N^{(\mathrm{Mag})}}\left(\frac{r}{\lambda^2}+\frac{\mu_0}{2}\coth\mu_0r-\frac{1}{2r}\right)(\hat{\boldsymbol{x}}\cdot\boldsymbol{\sigma}).\label{A_0 of HS1 in Magnetic}
\end{align}
We observe that the gauge connection \eqref{A_i of HS1 in Magnetic} and \eqref{A_0 of HS1 in Magnetic} is non-singular at the origin in this gauge.
The reason for the term ``magnetic" is that, in the large scale limit $\lambda\to\infty$, the gauge connection tends to,
\begin{align}\label{A_i of BPS1}
     A_j^{(\mathrm{Mag})}&\rightarrow -i\left(\frac{1}{2r}-\frac{\mu_0}{2\sinh\mu_0r}\right)(\hat{\boldsymbol{x}}\times \boldsymbol{\sigma})^j,\\
    A_0^{(\mathrm{Mag})}&\rightarrow -i\left(\frac{\mu_0}{2}\coth\mu_0r-\frac{1}{2r}\right)(\hat{\boldsymbol{x}}\cdot\boldsymbol{\sigma}),\label{A_0 of BPS1}
\end{align}
which is the well-known Bogomolnyi-Prasad-Sommerfield monopole gauge field of unit magnetic charge.
Note again that HS1 in this gauge \eqref{A_i of HS1 in Magnetic} and \eqref{A_0 of HS1 in Magnetic} is periodic in $x_0$ with period $2\pi/\mu_0$, and that its large-scale, or monopole, limit has translational symmetry in $x_0$-direction.
Therefore, the clutching function in the magnetic gauge also takes a value in the center of $\mathrm{SU}(2)$.
On the other hand, the relative sign of the clutching functions differs between the Lorenz --'t Hooft gauge and the magnetic gauge, as shown in Appendix \ref{App: clut}, so the corresponding gauge transformation swaps the element of the center. 
Such a gauge transformation for HS1 can be found easily in terms of the Weyl spinor \eqref{Weyl spinor} of the Nahm construction.
The boundary component $U$ \eqref{Weyl boundary component of Mag} is a quaternion consisting entirely of imaginary terms, and it can be converted into a real quaternion by multiplying a unit quaternion
\begin{align}\label{From Mag to LtH}
    q(x):=\exp \big(i\theta(x_0,r)(\hat{\boldsymbol{x}}\cdot\boldsymbol{\sigma})\big)=-\frac{\mathcal{Q}^\dagger}{|\mathcal{Q}|}=-|\mathcal{Q}|\mathcal{Q}^{-1},
\end{align}
from the right of the spinor \eqref{Weyl of mag}.
Here, the function $\theta(x_0,r)$ is 
\begin{align}\label{theta: exponent of q(x)}
    \theta(x_0,r)=\frac{\pi}{2}+\tan^{-1}\left(\tan\frac{\mu_0x_0}{2}\coth\frac{\mu_0r}{2}\right),
\end{align}
where the argument of arctangent is chosen such that the gauge element is continuous in the patch 1 of \eqref{patches of S^1}.
The multi-valuedness of arctangent does not matter for the function of the gauge transformation, and we take one of its branches here.
We will discuss the effect of this multi-valuedness of arctangent later.
The new spinor, denoted $V^{\mathrm{(LtH)}}$, takes the form,
\begin{align}\label{Weyl of LtH}
   V^{\mathrm{(LtH)}}:=V^{\mathrm{(Mag)}}q=\sqrt{\frac{r}{N^{\mathrm{(Mag)}}\sinh\mu_0r}}\begin{pmatrix}
        U\\
        \boldsymbol{v}(s)
    \end{pmatrix}q\nonumber\\
    =\frac{1}{\sqrt{N^{\mathrm{(LtH)}}}}\begin{pmatrix}
        1\\
        -\frac{\sqrt{\mu_0}\lambda}{2}\boldsymbol{v}(s)\mathcal{Q}^{-1}
    \end{pmatrix},
\end{align}
where $N^{\mathrm{(LtH)}}$ is defined in \eqref{N of LtH}.
Note that a unit quaternion is identified with an element of the gauge group SU(2), so the multiplication by $q$ \eqref{From Mag to LtH} induces a gauge transformation of the gauge connection.
Indeed, we can confirm the Weyl spinor \eqref{Weyl of LtH} generates the gauge connection of HS1 in the Lorenz --'t Hooft gauge \eqref{A_i of HS1 in LtH} and \eqref{A_0 of HS1 in LtH}. 
This means that the gauge transformation \eqref{From Mag to LtH}, and its inverse, produce or eliminate the singularity of the gauge connection at the origin.

For calorons of Harrington-Shepard type with $k$-singularities, i.e., the HS$k$-calorons in the Lorenz --'t Hooft gauge, the Weyl spinors generally take the form,
\begin{align}\label{Weyl for HSk}
    V=\frac{1}{\sqrt{N}}
    \begin{pmatrix}
        1\\
        \boldsymbol{v}_1(s)\mathcal{Q}_1\\
        \vdots\\
        \boldsymbol{v}_k(s)\mathcal{Q}_k
    \end{pmatrix},
\end{align}
where $\mathcal{Q}_j,\;(j=1,\dots,k)$ are certain quaternions similar to the bulk component of Lorenz --'t Hooft gauge \eqref{Weyl of LtH} and 
\begin{align}
    \boldsymbol{v}_k(s)=e^{i\boldsymbol{x}_j^\dagger s}=e^{i(x_0-a_{j,0})s}(\cosh r_js-\sinh r_js \;\hat{\boldsymbol{x}}_j\cdot\boldsymbol{\sigma}),
\end{align}
with $r_j=|\vec{x}-\vec{a}_j|$, the position of singularities denoted $a_j:=(a_{j,0},\vec{a}_j)$,  $\hat{\boldsymbol{x}}_j$ the unit vector in the direction of $\vec{x}-\vec{a}_j$, and $N$ a normalization factor.
We will consider these kinds of calorons with multi-singularities in a forthcoming paper.

\subsubsection{HS1 caloron in temporal gauge}
Another gauge of importance in the present study is the $A_0=0$ gauge of HS1 \cite{Dunne:2000if}, often referred to as the temporal gauge or the Weyl gauge, labelled (Temp) here.
The gauge transformation from the magnetic gauge to the temporal gauge is given by
\begin{align}\label{from Mag to Temp}
    G(x_0,r)=\exp(iu(x_0,r)\hat{\boldsymbol{x}}\cdot\boldsymbol{\sigma}),
\end{align}
where
\begin{align}\label{u of G}
    u(x_0,r)=p(r)\left\{\frac{\pi}{2}+\tan^{-1}\left(s(r)\tan\frac{\mu_0 x_0}{2}\right)\right\},
\end{align}
with
\begin{align}\label{def. p(r)}
    p(r)=\frac{(2r^2-\lambda^2)\sinh\mu_0 r+\mu_0\lambda^2r\cosh\mu_0 r}{r\sqrt{\sinh\mu_0 r\{4\mu_0\lambda^2 r\cosh\mu_0 r+(4r^2+\mu_0^2\lambda^4)\sinh\mu_0 r\}}},
\end{align}
and
\begin{align}
    s(r)=\frac{\mu_0\lambda^2\sinh\mu_0 r+2r(1+\cosh\mu_0 r)}{\sqrt{\sinh\mu_0 r\{4\mu_0\lambda^2 r\cosh\mu_0 r+(4r^2+\mu_0^2\lambda^4)\sinh\mu_0 r\}}}.\label{def. s(r)}
\end{align}
The branch of the arctangent in \eqref{u of G} is taken to be zero, and the gauge group element $G(x_0,r)$ is chosen such that it becomes the identity element of SU(2) at $x_0=-\beta/2=-\pi/\mu_0$ here. 
We then find that 
\begin{align}\label{A_0 in Temp}
    A_0^{\mathrm{(Temp)}}=G^{-1}A_0^{\mathrm{(Mag)}}G+G^{-1}\partial_0 G=0,
\end{align}
while the $\mathbb{R}^3$-components
\begin{align}\label{A_j in Temp}
    A_j^{\mathrm{(Temp)}}=G^{-1}A_j^{\mathrm{(Mag)}}G+G^{-1}\partial_j G,
\end{align}
take a somewhat complicated form.
Note that the gauge group element \eqref{from Mag to Temp} has no singularity, in particular at the origin, where $p(r)\sim O(r)$.
Thus, the gauge connection in the temporal gauge \eqref{A_j in Temp} is also non-singular everywhere, since the gauge connection in the magnetic gauge is likewise non-singular.
The most distinctive property of the temporal gauge is the presence of a non-trivial clutching function \cite{Dunne:2000if},
\begin{align}\label{clut for Temp}
    c(\vec{x})=c(r)=\exp(-i\pi p(r)\,\hat{\boldsymbol{x}}\cdot\boldsymbol{\sigma}),
\end{align}
where $p(r)$ is defined in \eqref{def. p(r)}.
The derivation of the gauge transformation \eqref{from Mag to Temp} and the clutching function \eqref{clut for Temp} is given in Appendix \ref{App: clut}.
We observe that $c(r)$ tends to a constant at the spatial infinity, $c(r)\to -1$ as $r\to\infty$, which can be seen from the asymptotic behaviour of $p(r)$ \eqref{def. p(r)}, namely $p(r)\xrightarrow[]{}1$ as $r\to\infty$, consistent with the assumption made in the definition of the clutching functions.

\subsection{Analytic evaluation of the second Chern number}
Having obtained the analytic forms of the HS1 gauge connection in several gauges, we now calculate the topological character of this gauge configuration analytically in each gauge.

\subsubsection{The Lorenz --'t Hooft gauge}
The gauge connection of the HS1 caloron in the Lorenz --'t Hooft gauge is given in \eqref{A_i of HS1 in LtH} and \eqref{A_0 of HS1 in LtH}, which has exact periodicity along $S^1$.  
Therefore, the clutching function in this gauge belongs to the center of SU(2) everywhere; consequently, the C-integral \eqref{Clut integral result} is trivially evaluated to zero.
The contribution to the second Chern number comes from the K-integral \eqref{K integral result}.

First, we evaluate the FA-part \eqref{FA-part} of \eqref{K integral result}, which is expressed in components as
\begin{align}\label{FA-part in components}
   -\frac{1}{8\pi^2} \int_{S^1\times(\partial\mathbb{R}^3)}2\tr F\wedge A_0=
    -\frac{1}{8\pi^2}\int_{S^1}dx_0\int_{\partial\mathbb{R}^3}r^2d\Omega\;\varepsilon_{ijk}\tr F_{ij}A_0 n_k,
\end{align}
where $\partial\mathbb{R}^3$ is a boundary two-sphere of radius $r$, and $d\Omega$ and $n_k$ are the solid angle element and the component of its unit normal vector $\boldsymbol{n}$, respectively.
The possible boundaries of $\mathbb{R}^3$ in this gauge are composed of the two-sphere at infinity $S^2_\infty$ and the two-sphere surrounding the origin $S_\varepsilon^2$ with infinitesimal radius $\varepsilon$, since the gauge connection is singular at the origin. 
The unit normal vectors $\boldsymbol{n}$ of $S^2_\infty$ and $S^2_\varepsilon$ are defined as outward and inward, i.e., $\pm(\sin\vartheta\cos\varphi,\sin\vartheta\sin\varphi,\cos\vartheta)$ in the polar coordinates on $\mathbb{R}^3$, respectively. 
The integrand is calculated straightforwardly from \eqref{A_i of HS1 in LtH} and \eqref{A_0 of HS1 in LtH} as,
\begin{align}\label{FA-integrand of LtH}
    \varepsilon_{ijk}\tr F^{\mathrm{(LtH)}}_{ij}&A^{\mathrm{(LtH)}}_0 n_k
    =\mp\frac{\mu_0^3\lambda^6\left\{\mu_0 r(\cos\mu_0 x_0\cosh\mu_0 r-1)+\sinh\mu_0 r(\cosh\mu_0 r-\cos\mu_0 x_0)\right\}}{2r^3(\cos\mu_0x_0-\cosh\mu_0 r)(2r(\cos\mu_0 r-\cosh\mu_0 r)-\mu_0\lambda^2\sinh\mu_0r)^3}\nonumber\\
    &\times\left\{1-\cosh2\mu_0 r+\frac{8r\cos\mu_0 x_0}{\mu_0\lambda^2}(\sinh\mu_0 r-\mu_0 r\cosh\mu_0 r)+2\mu_0^2r^2+\frac{4r^2}{\lambda^2}\left(2-\frac{\sinh2\mu_0 r}{\mu_0r} \right)\right\},
\end{align}
where the overall sign corresponds to the direction of the normal vector $\boldsymbol{n}$: minus for the outward case and and plus for the inward case.
The behaviour of the right-hand side as $r\to\infty$ is $O(1/r^5)$, so the integrand together with the surface element $r^2d\Omega$ tends to zero; thus, there is no contribution from $S^2_\infty$.
On the other hand, the behaviour of the integrand near the origin is
\begin{align}\label{FA-integrand at the origin}
   \lim_{r\to\varepsilon} \varepsilon_{ijk}\tr F^{\mathrm{(LtH)}}_{ij}A^{\mathrm{(LtH)}}_0 n_k=-\frac{\mu_0^8\lambda^4(2+\cos\mu_0x_0)(8+\mu_0^2\lambda^2+4\cos\mu_0x_0)}{9(2+\mu_0^2\lambda^2-2\cos\mu_0x_0)^3(\cos\mu_0x_0-1)}\varepsilon+O(\varepsilon^3), 
\end{align}
where the plus sign in \eqref{FA-integrand of LtH} applies for the boundary $S^2_\varepsilon$.
The integrand is therefore non-singular at the origin, despite the gauge connection itself being singular there.
Consequently, it is not necessary to consider the surface integral over $S^2_\varepsilon$, and we find that there is no contribution from \eqref{FA-part in components} in the Lorenz --'t Hooft gauge. 
Furthermore, in the large-scale, or monopole, limit, the integrand of \eqref{FA-integrand of LtH} becomes,
\begin{align}\label{FA-integrand of LtH at large scale}
    &\varepsilon_{ijk}\tr F^{\mathrm{(LtH)}}_{ij}A^{\mathrm{(LtH)}}_0 n_k
    \xrightarrow[\lambda\to\infty]{}\nonumber\\
    &\pm\frac{\left\{\mu_0 r(\cos\mu_0 x_0\cosh\mu_0 r-1)+\sinh\mu_0 r(\cosh\mu_0 r-\cos\mu_0 x_0)\right\}(1+2\mu_0^2r^2-\cosh2\mu_0r)}{2r^3\sinh^3\mu_0 r(\cosh\mu_0r-\cos\mu_0x_0)},
\end{align}
which behaves as $O(1/r^3)$ as $r\to\infty$.
Despite this slower asymptotic decay compared to the finite $\lambda$ case, the integral \eqref{FA-integrand of LtH at large scale} over $S^2_\infty$ also vanishes.
Hence, there is no magnetic charge in this gauge even in the large-scale limit $\lambda\to\infty$. 
We conclude that there is no contribution to the FA-part \eqref{FA-part} of the K-integral \eqref{K integral} with both finite and infinite $\lambda$ cases.

Next, we calculate the dAA-part \eqref{dAA-part} of the K-integral \eqref{K integral}, expressed in components as,
\begin{align}\label{dAA in components}
    -\frac{1}{8\pi^2}\int_{S^1\times(\partial\mathbb{R}^3)}\tr \partial_0A\wedge A=-\frac{1}{8\pi^2}\int_{S^1}dx_0\int_{\partial\mathbb{R}^3}r^2d\Omega\;
    \varepsilon_{ijk}\tr (\partial_0A_i)A_j n_k.
\end{align}
On the boundary two-sphere $\partial\mathbb{R}^3$ of radius $r$, the integrand is 
\begin{align}\label{dAA integrand LtH}
    \varepsilon_{ijk}\tr (\partial_0A^{\mathrm{(LtH)}}_i)A^{\mathrm{(LtH)}}_j n_k=
    \mp\frac{\mu_0^4\lambda^4\sinh\mu_0r\left\{\mu_0 (\cosh\mu_0 r-\cos\mu_0x_0)-\frac{\sinh\mu_0 r}{r}(1-\cos\mu_0 x_0\cosh\mu_0r)\right\}}{(2r)^2(\cosh\mu_0r-\cos\mu_0x_0)^2\left(\cosh\mu_0r-\cos\mu_0 x_0+\frac{\mu_0\lambda\sinh\mu_0 r}{2r}\right)^2},
\end{align}
where the minus sign corresponds to the outward normal vector, and the plus sign to the inward one.
The behaviour of \eqref{dAA integrand LtH} as $r\to \infty$ is $e^{-\mu_0r}/r^3$, so the boundary integral over $S^2_\infty$ vanishes, and there is no contribution for \eqref{dAA in components} from the boundary at infinity. 
We now consider the contribution from the infinitesimal boundary sphere $S^2_\varepsilon$ surrounding the origin of $\mathbb{R}^3$.
From \eqref{dAA integrand LtH}, we extract the relevant part of the integrand multiplied by the surface factor $r^2$ at $r=\varepsilon$,
\begin{align}
    r^2\varepsilon_{ijk}\tr (\partial_0A^{\mathrm{(LtH)}}_i)A^{\mathrm{(LtH)}}_j n_k \Big|_{r=\varepsilon}
    =\frac{\mu_0^4\lambda^2}{4}\left\{\frac{\mu_0\sinh\mu_0\varepsilon}{\cosh\mu_0\varepsilon-\cos\mu_0x_0}-\frac{\sinh^2\mu_0\varepsilon}{\varepsilon}\frac{1-\cos\mu_0x_0\cosh\mu_0\varepsilon}{(\cosh\mu_0\varepsilon-\cos\mu_0x_0)^2}\right\}\nonumber\\
    \times\left(\cosh\mu_0\varepsilon-\cos\mu_0x_0+\frac{\mu_0\lambda\sinh\mu_0 \varepsilon}{2\varepsilon}\right)^{-2},
\end{align}
where the inverse square factor is regular as $\varepsilon\to0$ and at $x_0=0$, the first term in the bracket is singular, and the second term of that is $O(\varepsilon)$.
As we will show in Appendix \ref{App: delta}, the singular behaviour of the first term is encoded in the periodic delta function,
\begin{align}\label{periodic delta}
\lim_{\varepsilon\to0}\frac{\mu_0\sinh\mu_0\varepsilon}{\cosh\mu_0\varepsilon-\cos\mu_0x_0}=2\pi\sum_n\delta\left(x_0-\frac{2\pi n}{\mu_0}\right),
\end{align}
in the language of Sato's hyperfunction theory \cite{imai1992applied,morimoto1993introduction}.
Hence, the regularization of the singular point of $S^1$ at $x_0=0$ is automatically incorporated, and the evaluation of \eqref{dAA in components} over $S_\varepsilon^2$ gives
\begin{align}\label{dAA charge in LtH}
    &\lim_{\varepsilon\to0}\frac{-1}{8\pi^2}\int_{-\pi/\mu_0}^{\pi/\mu_0}dx_0\int_{S^2_\varepsilon}\varepsilon_{ijk}\tr (\partial_0A^{\mathrm{(LtH)}}_i)A^{\mathrm{(LtH)}}_j n_k\varepsilon^2d\Omega\nonumber\\
    =&\frac{-1}{8\pi^2}(4\pi)\int_{-\pi/\mu_0}^{\pi/\mu_0}dx_0\frac{\mu_0^4\lambda^2}{4}2\pi\delta(x_0)\left(1-\cos\mu_0x_0+\frac{\mu_0^2\lambda}{2}\right)^{-2}=-1,
\end{align}
as expected.
Note that this evaluation remains valid even in the large-scale limit $\lambda\to\infty$.
To summarize this subsection, we have found that the second Chern number $Q$ \eqref{2nd Chern number} evaluated using the Lorenz --'t Hooft gauge, equal minus unity, arising entirely from the dAA-part \eqref{dAA-part} of the K-integral \eqref{K integral}.
The topological charge is generated by the singularity of the gauge connection at the origin, which manifests through the delta function in this gauge.
The FA-part \eqref{FA-part} does not contribute even at the infinite $\lambda$ case.

\subsubsection{The magnetic gauge}
Consider next the HS1 caloron in the magnetic gauge \eqref{A_i of HS1 in Magnetic} and \eqref{A_0 of HS1 in Magnetic} for the evaluation of the second Chern number.
The contribution of the C-integral \eqref{Clut integral} is trivially zero, since the clutching function lies in the center of SU(2), as in the Lorenz --'t Hooft gauge.
We therefore focus again on the K-integral \eqref{K integral result}.

In contrast to the Lorenz --'t Hooft gauge, the gauge connection is regular at the origin in this gauge.
Hence, the only spatial boundary to be considered is the two-sphere at infinity $S^2_\infty$.
For the FA-part \eqref{FA-part} of the K-integral, the gauge connection \eqref{A_i of HS1 in Magnetic} and \eqref{A_0 of HS1 in Magnetic} gives the integrand of \eqref{FA-part in components},
\begin{align}\label{FA integrand in Mag}
\varepsilon_{ijk}\tr &F^{(\mathrm{Mag})}_{ij}A^{(\mathrm{Mag})}_0 n_k
    =-\frac{\left(\frac{2r^2}{\lambda^2}\sinh\mu_0r+\mu_0r\cosh\mu_0r-\sinh\mu_0r\right)}
    {2r^3\left(\sinh\mu_0r+\frac{2r}{\mu_0\lambda^2}(\cosh\mu_0r-\cos\mu_0x_0)\right)^3}\nonumber\\
    &\times \left\{1-\cosh2\mu_0r+2\mu_0^2r^2+\frac{4r}{\mu_0\lambda^2}\Bigl(2\mu_0r-2\cos\mu_0x_0(\mu_0r\cosh\mu_0r-\sinh\mu_0r)-\sinh2\mu_0r\Bigr)\right\},
\end{align}
where the normal vector of the boundary two-sphere is taken to be outward.
As $r\to\infty$, the behaviour of \eqref{FA integrand in Mag} is $O(1/r^3)$, which decays more slowly than in the Lorenz --'t Hooft gauge; nevertheless, the integral \eqref{FA-part in components} over $S^2_\infty$ evaluates to zero similarly.
Hence, the FA-part \eqref{FA-part in components} gives no contribution for finite $\lambda$.
On the other hand, the situation changes qualitatively in the large-scale, or monopole, limit $\lambda\to\infty$.
As observed in the Lorenz --'t Hooft gauge, the asymptotic decay of the FA-part integrand becomes slower in this limit. 
In the present gauge, the integrand of \eqref{FA integrand in Mag} becomes,
\begin{align}
   \lim_{\lambda\to\infty}\varepsilon_{ijk}\tr &F^{(\mathrm{Mag})}_{ij}A^{(\mathrm{Mag})}_0 n_k
   =\frac{\mu_0}{r^2},
\end{align}
so that the FA-integral over $S^2_\infty$ yields a non-zero value,
\begin{align}\label{Monopole charge in Mag}
    -\frac{1}{8\pi^2}\int_{-\pi/\mu_0}^{\pi/\mu_0}dx_0\int_{\partial\mathbb{R}^3}r^2d\Omega\;\varepsilon_{ijk}\tr F^{(\mathrm{Mag})}_{ij}A^{(\mathrm{Mag})}_0 n_k \xrightarrow[\lambda\to\infty]{}
    -\frac{1}{8\pi^2}\int_{-\pi/\mu_0}^{\pi/\mu_0}dx_0\int_{S^2_\infty}d\Omega\;(-\mu_0)=-1,
\end{align}
which is interpreted as a unit magnetic charge.
The result is consistent with the BPS monopole limit \eqref{A_i of BPS1} and \eqref{A_0 of BPS1} of the magnetic gauge.
Hence, the FA-part \eqref{FA-part} contributes only in the large-scale limit in the magnetic gauge. 

Consider next the dAA-part, for which the integrand of \eqref{dAA in components} in this gauge is
\begin{align}\label{dAA integrand in Mag}
    \varepsilon_{ijk}&\tr (\partial_0A^{(\mathrm{Mag})}_i)A^{(\mathrm{Mag})}_j n_k\nonumber\\
    =&\frac{2\mu_0}{r}\left\{-\mu_0\lambda^2\cos\mu_0x_0-2r(1+\cos\mu_0x_0)\coth\mu_0r+\frac{r(2+(2+\mu_0^2\lambda^2)\cos\mu_0x_0)}{\sinh\mu_0r}\right\}\nonumber\\
    &\times\left(\mu_0\lambda^2+2r\coth\mu_0r-\frac{2r\cos\mu_0x_0}{\sinh\mu_0r}\right)^{-2}.
\end{align}
Note that the integrand \eqref{dAA integrand in Mag} vanishes in the large-scale limit $\lambda$ as $O(1/\lambda^2)$.
On the other hand, the asymptotic form of \eqref{dAA integrand in Mag} for finite $\lambda$ as $r\to\infty$ is $\mu_0(1+\cos\mu_0x_0)/r^2$, so the dAA-part of this gauge evaluates to
\begin{align}
    -\frac{1}{8\pi^2}\int_{S^1\times S^2_\infty}\tr \partial_0A^{(\mathrm{Mag})}\wedge A^{(\mathrm{Mag})}=
    -\frac{1}{8\pi^2}\int_{-\pi/\mu_0}^{\pi/\mu_0}dx_0\int_{S^2_\infty}d\Omega (\mu_0(1+\cos\mu_0x_0))=-1,
\end{align}
contributing minus unity to the second Chern number $Q$.
The contributions to the K-integral of $Q$ are summarised in Table \ref{K-integral in Mag}.
The total value of the second Chern number remains unchanged under the large-scale limit.

    \begin{table}[H]
    \begin{center}
      \begin{tabular}{c|ccc|cccc}
         && finite $\lambda$ &&& $\lambda\to\infty$ \\
        \hline
        FA-part& &0&&&-1\\
        dAA-part& &-1&&&0
        \end{tabular}
    \caption{Contribution to $Q$ from the K-integral in the magnetic gauge}  
\label{K-integral in Mag}
    \end{center}
    \end{table}

\subsubsection{The temporal gauge}
In this subsection, we examine the second Chern number of HS1 in the temporal gauge given in \eqref{A_0 in Temp} and \eqref{A_j in Temp}.
The most significant difference from the previous cases is that the presence of a non-trivial clutching function \eqref{clut for Temp}, which yields a contribution to the first term of the C-integral \eqref{Clut integral result}.
Defining $c(r)=\exp(iw(r)(\hat{\boldsymbol{x}}\cdot\boldsymbol{\sigma}))$ and $c\partial_j c^{-1}=iW_{ja}\sigma_a$, a straightforward calculation shows that
\begin{align}\label{C from clutching function in temp}
    -\frac{1}{24\pi^2}\int_{S^3}\tr(c^{-1}dc)^3&=-\frac{1}{24\pi^2}\int_{S^3}i^3\varepsilon_{jkl}W_{ja}W_{kb}W_{lc}\tr(\sigma^a\sigma^b\sigma^c) d^3x\nonumber\\
    &=-\frac{1}{24\pi^2}\int_{S^3}12\,\mathrm{det}(W_{ja})d^3x\nonumber\\
    &=-\frac{1}{2\pi^2}\int_{S^2}d\Omega\int_0^\infty dr\, r^2\,\frac{-1}{r^2}\frac{dw}{dr}\sin^2w\nonumber\\
    &=\frac{2}{\pi}\int_{w(0)}^{w(\infty)}dw\sin^2 w=-1,
\end{align}
where we used $w(r)=-\pi p(r)$, $p(0)=0$, and $p(r)\to1$ as $r\to\infty$.
Hence, the C-integral for HS1 in the temporal gauge contributes minus unity.
The contribution is not affected by the large-scale limit $\lambda\to\infty$.
The second term of \eqref{Clut integral result} does not contribute, due to the asymptotic behaviour of the clutching function tending to a constant group element.
However, one cannot naively exclude the possibility of a non-zero contribution from this term; for example, if $w(r)\to -\pi/2$ together with $A_j\sim O(1/r)$ as $r\to\infty$, there would remain a non-trivial integral.

For the K-integral \eqref{K integral result}, there is no contribution in this gauge.
The FA-part \eqref{FA-part} is trivially zero due to the gauge condition $A_0^{\mathrm{(Temp)}}=0$.
For the evaluation of the dAA-part \eqref{dAA-part}, the only possible boundary of $\mathbb{R}^3$ is $S^2_\infty$.
The integrand of the dAA-part in the component form \eqref{dAA in components} can be expressed in terms of the gauge connection of the magnetic gauge \eqref{A_i of HS1 in Magnetic} and \eqref{A_0 of HS1 in Magnetic}, its field strength, and the gauge transformation \eqref{from Mag to Temp}, via \eqref{A_j in Temp}, as follows:
\begin{align}\label{dAA integrand in Temp}
    \varepsilon_{ijk}\tr (\partial_0A_i^{\mathrm{(Temp)}})A_j^{\mathrm{(Temp)}}n_k&=
    \varepsilon_{ijk}\tr (F_{0i}^{\mathrm{(Temp)}})A_j^{\mathrm{(Temp)}}n_k\nonumber\\
    &=\varepsilon_{ijk}\tr G^{-1}F_{0i}^{\mathrm{(Mag)}}G(G^{-1}A_j^{\mathrm{(Mag)}}G+G^{-1}\partial_j G)n_k\nonumber\\
    &=\varepsilon_{ijk}\tr F_{0i}^{\mathrm{(Mag)}}\left(A_j^{\mathrm{(Mag)}}+(\partial_j G)G^{-1})\right)n_k\nonumber\\
    &=\tr F^{\mathrm{(Mag)}}_{ik}(A^{\mathrm{(Mag)}}_i+(\partial_i G)G^{-1})n_k,
\end{align}
where we used $F_{0j}^{(\mathrm{Temp})}=\partial_0A_j^{(\mathrm{Temp})}$, which follows from $A_0^{(\mathrm{Temp})}=0$, and the anti-selfdual equation \eqref{ASD}.
The asymptotic behaviour of \eqref{dAA integrand in Temp} as  $r\to\infty$ is $O(1/r^4)$, and we find the dAA-part over the boundary $S^2_\infty$ vanishes in this gauge.
In the large scale limit $\lambda\to\infty$, the integrand \eqref{dAA integrand in Temp} behaves as $e^{-\mu_0r}/r$ at $S^2_\infty$, so there is likewise no contribution from the dAA-part in this limit.

This analytic calculation of the second Chern number $Q$ of HS1 in the temporal gauge is consistent with the numerical calculation previously reported in \cite{Dunne:2000if}.
We therefore conclude that $Q$ for HS1 in the temporal gauge equals minus one, as expected.
In contrast to the other gauges considered above,
the contribution arises from the C-integral \eqref{Clut integral}.

\subsection{Further consideration on various gauges}

Throughout this section, we have demonstrated the calculation of the second Chern number $Q$ of the HS1 caloron in three gauges: the Lorenz --'t Hooft gauge, the magnetic gauge, and the temporal gauge.
In all cases, they yield the same result $Q=-1$ as expected; however, the origin of the topological contribution depends on the choice of gauges. 

In the Lorenz --'t Hooft gauge, only the dAA-part \eqref{dAA-part} of the K-integral \eqref{K integral}, evaluated on the small boundary sphere $S^2_\varepsilon$ surrounding the origin, contributes, and does so regardless of the value of the scale parameter $\lambda$.
The contribution arises from a delta function singularity of the integrand. 
In contrast, neither the dAA-part nor the FA-part receives any contribution at spatial infinity $S^2_\infty$.
Hence, the HS1 caloron in the Lorenz --'t Hooft gauge carries no magnetic charge, even in the large scale limit.
The C-integral yields no contribution due to its trivial clutching function.

In the magnetic gauge, on the other hand, the integrands of dAA-part and FA-part are regular at the origin, so $S^2_\epsilon$ does not appear as a boundary of the K-integral.
For finite $\lambda$, the contribution to the dAA-part comes from the integral over $S^2_\infty$, while the FA-part contributes only in the limit $\lambda\to\infty$, where it is interpreted as a magnetic charge.
This suggests that the magnetic charge of calorons should be regarded as a gauge-dependent quantity.
We also observe that the gauge transformation between the Lorenz --'t Hooft gauge and the magnetic gauge \eqref{From Mag to LtH} removes or attaches the delta function singularity of the dAA-part at the origin. 

In contrast to these two gauges, the temporal gauge has a non-trivial clutching function \eqref{clut for Temp}, and the contribution of minus unity comes from the C-integral alone, irrespective of the value of $\lambda$.

The contributions to $Q$ in each gauge are listed in Table \ref{contribution to Q}.

    \begin{table}[H]
    \begin{center}
      \begin{tabular}{l||ccc|cccccccc}
         &\multicolumn{2}{c}{C-integral} && \multicolumn{6}{c}{K-integral}\\ \cline{5-12}
         &&&&&&&&&&&\\
       &&&&&  FA-part          && \multicolumn{3}{c}{dAA-part}\\
       &&$\mathbb{R}^3\simeq S^3$&&&  $S^2_\infty$     && $S^2_\varepsilon$ && $S^2_\infty$&\\
        \hline
         &&&&&&&&&&&\\
         LtH& &0&&&0&&-1&&0&&\\
        LtH$|_{\lambda\to\infty}$ & &0&&&0&&-1&&0&\\
        &&&&&&&&&&&\\
        Mag& &0&&&0&&-&&-1\\
        Mag$|_{\lambda\to\infty}$& &0&&&-1&&-&&0\\
        &&&&&&&&&&&\\
        Temp& &-1&&&0&&-&&0\\
        Temp$|_{\lambda\to\infty}$& &-1&&&0&&-&&0\\
        \end{tabular}
    \caption{Contribution to $Q$ in several gauges (the first column), the integration regions are indicated in the third row.}  
\label{contribution to Q}
    \end{center}
    \end{table}

As we have shown, the gauge transformation \eqref{From Mag to LtH} for HS1 in the magnetic gauge attaches the delta function singularity at the origin and removes the contribution to the K-integral from $S^2_\infty$.
This process can be iterated, shifting the origin of the contribution to the K-integral repeatedly.
Let us apply the gauge transformation \eqref{From Mag to LtH} to the gauge connection in the magnetic gauge, with the exponent \eqref{theta: exponent of q(x)} replaced by, 
\begin{align}\label{center transform m exponent}
      \theta_m(x_0,r)=m\qty{\frac{\pi}{2}+\tan^{-1}\left(\tan\frac{\mu_0x_0}{2}\coth\frac{\mu_0r}{2}\right)}, \ m\in\mathbb{Z},
\end{align}
namely,
\begin{align}\label{center transform of m}
    q_m:=\exp\left(i\theta_m(x_0,r)(\hat{\boldsymbol{x}}\cdot\boldsymbol{\sigma})\right).
\end{align}
In the transformed gauge, the contributions to the dAA-part of the K-integral from $S_\infty^2$ and $S^2_\varepsilon$ become $m-1$ and $-m$, respectively. 
The total value of the K-integral, namely minus unity, remains unchanged after this gauge transformation.
Note that, in the large-scale limit, the contribution to the FA-part from $S_\infty^2$ replaces that of the dAA-part; consequently, the magnetic charge is no longer restricted to minus unity, yet the total $Q=-1$ is preserved.
As we will see in Appendix \ref{App: clut}, the gauge transformation from the magnetic gauge to the Lorenz --'t Hooft gauge, i.e., the $m=1$ case, swaps the sign of the clutching functions.
The generalised transformation \eqref{center transform of m} likewise exchanges the sign as the branch $m$ shifts $\pm1$.
Since this sign difference corresponds to the center element of SU(2), we refer to the gauge transformation \eqref{center transform of m} as the center transformation.

Next, consider the generalization of the gauge transformation between the magnetic gauge and the temporal gauge, given in \eqref{from Mag to Temp}.
As we have seen, this transformation shifts the contribution to $Q$ from the K-integral to the C-integral.
In a similar fashion to the center transformation, it is possible to formulate a gauge transformation that exchanges the contribution to $Q$ between the C-integral and the K-integral by choosing the branch of arctangent in \eqref{from Mag to Temp}, namely
\begin{align}\label{medium transformation}
    G_n(x_0,r)=
    \exp(iu_n(x_0,r)\hat{\boldsymbol{x}}\cdot\boldsymbol{\sigma}),
\end{align}
with
\begin{align}\label{exponent in medium transformation}
     u_n(x_0,r)=-np(r)\left\{(n+2)\frac{\pi}{2}+\tan^{-1}\left(s(r)\tan\frac{\mu_0 x_0}{2}\right)\right\},\ n\in \mathbb{Z}.
\end{align}
The gauge transformation \eqref{medium transformation} shifts the contribution to $Q$ from the C-integral and the K-integral by $n-1$ and $-n$, respectively, while of course preserving the total value $Q=-1$.
Note that the transformation \eqref{medium transformation} acts non-trivially on both the internal structure of the K-integral and the behaviour of the clutching function; these aspects will be considered in a forthcoming paper.

\section{Concluding remarks}
In this paper, we have re-examined the topological number of calorons -- the instantons of pure Yang-Mills theory on $S^1\times \mathbb{R}^3$.
Through an analytic calculation of the second Chern number $Q$ \eqref{2nd Chern number} for the Harrington-Shepard caloron of $Q=-1$ explicitly, we find the Nye-Singer formula for $Q$ \cite{Nye:2000eg} requires some modification, due to an additional contribution from the dAA-part integral \eqref{dAA-part}, such as
\begin{align}\label{Q modified formula}
     Q&=-\frac{1}{24\pi^2}\int_{\mathbb{R}^3\simeq S^3}\tr(dc\,c^{-1})^3-\frac{1}{8\pi^2}\int_{S^1\times S^2_\infty}\tr 2F\wedge A_0-\frac{1}{8\pi^2}\int_{S^1\times(\partial\mathbb{R}^3)}\tr \partial_0A\wedge A.
\end{align}
In addition, as we have mentioned in the illustration of the temporal gauge, there could be a contribution from the second term of \eqref{Clut integral result}, although there is no explicit example of the case at present.
The consideration in \cite{Nye:2000eg} assumed the smoothness of field configurations over the whole space, meaning that a special gauge condition was implicitly imposed; however, a general formula should apply to any gauge.
In fact, we have shown that the calculation of $Q$ is executable in the case of the singular gauge, such as the Lorenz --'t Hooft gauge.

To establish the complete formula of the second Chern number of calorons, we need to investigate $Q$ for more general types of configurations.  
One such case is the Harrington-Shepard calorons with multiple singularities, HS$k$, for which the singularity structure would be complicated.
In contrast to the HS1 case considered in this paper, there would not exist a gauge transformation that removes all of the singularities in $\mathbb{R}^3$.
In fact, we find from the Weyl spinor structure of the Nahm construction \eqref{Weyl for HSk} that not all of the singularities in $\mathbb{R}^3$ can be removed by a simple gauge transformation. 
Hence, the calculation of $Q$ for HS$k$ requires careful treatment of these singularities.

Another class consists of calorons with non-trivial holonomy parameters, such as those in \cite{Cork:2017hnj,Kraan:1998pm,Lee:1998bb,Bruckmann:2004nu,Harland:2007zz,Nakamula:2009sq,Kato:2018zfb}.
A characteristic feature of these calorons is their constituent monopole structure \cite{Foscolo:2022tas}. 
One expects a contribution to $Q$ from the second term of \eqref{Nye-Singer formula} in such configurations, irrespective of the value of the scale parameter, which would generally result in a non-integer value of $Q$.
A point that should be clarified is under what conditions 
$Q$ takes integer values for such calorons.
In addition, as we have observed, the dAA-part of the K-integral always yields integer values in several gauges; however, the reason for this remains unclear, in contrast to the integer valuedness of the C-integral, which is an element of the homotopy group $\pi_3(S^3)$.
These results will be addressed in a forthcoming paper. 

As we have seen, the non-trivial topology of the $S^1$- compactified space leads to a complicated structure of the topological invariant of field configurations.
In the context of research on the quark confinement mechanism of Yang-Mills theory, studies are actively progressing for theories on the other $S^1$-compactified base spaces, such as $T^2\times\mathbb{R}^2, \, T^3\times\mathbb{R}$ and $T^4$, as mentioned in the Introduction.
Further examination of the topological invariant on such non-simply connected base spaces will be a crucial issue for those studies.

\section*{Acknowledgement}
The authors wish to thank Krishna Kalluri for his careful reading of the rivised manuscript and useful commnents.
AN is supported in part by JSPS KAKENHI Grant Number JP23K02794.
He also thanks all the participants of Soliton Break Through Workshop 2025 (SBREW 2025) for fruitful discussions, particularly the organiser, Filip Blashke, for his kind hospitality. 

\section*{Conflict of interest}
The authors have no conflicts to disclose.

\appendix
\renewcommand{\theequation}{\Alph{section}\arabic{equation}}
\setcounter{equation}{0}

\section{Clutching functions and gauge transformations}\label{App: clut}
As defined in \eqref{clutching function i} and \eqref{clutching function 0} of Section \ref{Sec: Topological Character}, the clutching functions relate the gauge connections at the endpoints $x_0=\pm \beta/2$ of a patch of $S^1$. 
Here, we derive the clutching function for the temporal gauge via the gauge transformation from the magnetic gauge, whose clutching function is trivial.  

Consider a gauge transformation from a first gauge $A_j$ to a second gauge $A'_j$, with clutching functions $c(\vec{x})$ and $c'(\vec{x})$, respectively, defined by
\begin{align}\label{clut for 1st gauge}
    A_j(\beta/2,\vec{x})&=c(\vec{x})A_j(-\beta/2,\vec{x})c^{-1}(\vec{x})+c(\vec{x})\partial_j c^{-1}(\vec{x}),\\
    A'_j(\beta/2,\vec{x})&=c'(\vec{x})A'_j(-\beta/2,\vec{x})c'^{-1}(\vec{x})+c'(\vec{x})\partial_j c'^{-1}(\vec{x}).\label{clut for 2nd gauge}
\end{align}
The temporal components of gauge connections are not relevant here.
We now show that the clutching functions of both gauges are related by
\begin{align}\label{gauge transform for clut}
    c'(\vec{x})=g(\beta/2,\vec{x})\;c(\vec{x})\;g^{-1}(-\beta/2,\vec{x}),
\end{align}
where $g(x_0,\vec{x})$ is the gauge transformation from the first gauge to the second gauge. 

The gauge connections of the second gauge at $x_0=\beta/2$ can be rewritten in terms of that at $-\beta/2$ as 
\begin{align}\label{link of A and A'}
    A'_j(\beta/2)&=g(\beta/2)A_j(\beta/2)g^{-1}(\beta/2)+g(\beta/2)\partial_jg^{-1}(\beta/2)\nonumber\\
    &=g(\beta/2)\{c A_j(-\beta/2)c^{-1}+c\partial_jc^{-1}\}g^{-1}(\beta/2)+g(\beta/2)\partial_jg^{-1}(\beta/2),
\end{align}
where we used \eqref{clut for 1st gauge} and suppressed the $\vec{x}$ dependence.
On the other hand, the left-hand side of \eqref{link of A and A'} must also satisfy 
\begin{align}\label{link of A and A' LHS}
     A'_j(\beta/2)&=c'A'_j(-\beta/2)c'^{-1}+c'\partial_j c'^{-1}\nonumber\\
    &=c'\left\{g(-\beta/2)A_j(-\beta/2)g^{-1}(-\beta/2)+g(-\beta/2)(\partial_jg^{-1}(-\beta/2))\right\}c'^{-1}+c'\partial c'^{-1}.
\end{align}
Consistency between \eqref{link of A and A'} and \eqref{link of A and A' LHS} requires $g(\beta/2)c=c'g(-\beta/2)$, which confirms the relation \eqref{gauge transform for clut}.

We now determine the clutching function of the temporal gauge \eqref{clut for Temp} from \eqref{gauge transform for clut}.
Taking the magnetic gauge as the first gauge and the temporal gauge as the second, the clutching functions are related via the gauge transformation $G(x_0,\vec{x})$ of \eqref{from Mag to Temp} as, 
\begin{align}
    c^{(\mathrm{Temp})}=G^{-1}(\beta/2)c^{(\mathrm{Mag})}G(-\beta/2).
\end{align}
The derivation of the gauge transformation \eqref{from Mag to Temp} is given below.
Since the clutching function of the magnetic gauge belongs to the center of SU(2), and the gauge transformation has been chosen such that $G(-\beta/2)=1$, we find from \eqref{from Mag to Temp} and \eqref{u of G}, using $\tan \pi/2\to+\infty$
\begin{align}\label{clut for temp from p(r)}
    c^{(\mathrm{Temp})}=\pm G^{-1}(\beta/2)=\pm \exp(-i\pi p(r)\hat{\boldsymbol{x}}\cdot\boldsymbol{\sigma}),
\end{align}
where $p(r)$ is defined in \eqref{def. p(r)}.
Note that the sign ambiguity in \eqref{clut for temp from p(r)} arises from the choice of $G(-\beta/2)$ 
and $c^{(\mathrm{Mag})}$; the plus sign is adopted in the main text.

Furthermore, we consider the relation between the clutching functions of the Lorenz --'t Hooft gauge and the magnetic gauge, both of which take values in the center of SU(2).
The gauge transformation $q(x_0)$ between these gauges is given in \eqref{From Mag to LtH}; we thus find,
\begin{align}\label{clut: swapping of sign}
    c^{(\mathrm{LtH})}=q^{-1}(\beta/2)c^{(\mathrm{Mag})}q(-\beta/2)=-c^{(\mathrm{Mag})},
\end{align}
where we used 
\begin{align}
    q(\pm \beta/2)=\sqrt{\frac{2}{\cosh\mu_0r+1}}\left(\mp\cosh\frac{\mu_0r}{2}\right)=\mp1.
\end{align}
We therefore find that the gauge transformation between these gauges swaps the clutching functions within the center of SU(2).

We now derive the gauge transformation \eqref{from Mag to Temp}.
From the definition of the temporal gauge condition \eqref{A_0 in Temp}, the differential equation for $G(x_0,\vec{x})$ is,
\begin{align}\label{Differential equation for G}
A_0^{\mathrm{(Temp)}}=G^{-1}A_0^{\mathrm{(Mag)}}G+G^{-1}\partial_0 G=0\nonumber\\
\Leftrightarrow \partial_0 G=-A_0^{\mathrm{(Mag)}}G,
\end{align}
where $A_0^{\mathrm{(Mag)}}$ is given in \eqref{A_0 of HS1 in Magnetic}.
This equation is solved by a path-ordered integral along $S^1$
\begin{align}
    G(x_0,\vec{x})=\mathcal{P}\exp\left(-\int_{-\beta/2}^{x_0}A_0^{\mathrm{(Mag)}}(x'_0,\vec{x})dx'_0\right),
\end{align}
with the boundary condition $G(-\beta/2,\vec{x})=1$.
Since $A_0^{\mathrm{(Mag)}}$ commutes with itself at different values of $x_0$, owing to the matrix structure, the path-ordered integral reduces to an ordinary integral, and we obtain \eqref{from Mag to Temp}.

\section{Periodic delta function}\label{App: delta}
Here we provide a justification for identifying \eqref{periodic delta} as the periodic delta function.
In Sato's hyperfunctions theory \cite{imai1992applied,morimoto1993introduction}, hyperfunctions, which includes the delta function as a special case, are defined as difference of boundary values of holomorphic functions in the upper-half-plane (UHP) and the lower-half-plane (LHP) of a complex plane.
These holomorphic functions are referred to as the defining functions of a hyperfunction.
We restrict ourselves here to one-variable case. 
As an illustration, consider the delta function with a support at the origin $x=0$.
The corresponding defining functions are
\begin{align}\label{defining functions}
    F_\pm(z):=-\frac{1}{2\pi i}\frac{1}{z\pm i\varepsilon},
\end{align}
where $z\in\mathbb{C}$, and $\varepsilon$ is understood to be infinitesimal.
The functions $F_+$ and $F_-$ are holomorphic at UHP and LHP, respectively, since their poles lie in the opposite half-planes.
The difference of the boundary values on $\mathbb{R}$ is
\begin{align}\label{boundary values}
   F_+(x)-F_-(x)= -\frac{1}{2\pi i}\left(\frac{1}{x+i\varepsilon}-\frac{1}{x-i\varepsilon}\right),
\end{align}
where $x\in\mathbb{R}$.
We next multiply \eqref{boundary values} by a test function $f(z)$ and integrate over $\mathbb{R}$, deforming the contour near the origin along a small clockwise circle $C$ to avoid the pole,
\begin{align}\label{integral on R}
    -\frac{1}{2\pi i}\int_{-\infty}^{\infty}\left(\frac{1}{x+i\varepsilon}-\frac{1}{x-i\varepsilon}\right)f(x)dx.
\end{align}
The contribution from the two terms in \eqref{integral on R} cancel almost everywhere on $\mathbb{R}$, leaving only the residue at the origin
\begin{align}
    -\frac{1}{2\pi i}\oint_C\frac{f(z)}{z}dz=f(0),
\end{align}
where Cauchy's integral formula has been applied.
We therefore conclude that \eqref{boundary values} can be identified with a delta function with support at the origin, in regularised form.

We now construct the delta function with period $\pi$.
From the formula,
\begin{align}\label{def. cotangent}
    \lim_{m\to\infty}\sum_{n=-m}^m\frac{1}{x+n\pi}=\cot x,
\end{align}
we define the defining functions
\begin{align}
    F_\pm(z)=\frac{i}{2\pi}\cot(z\pm i\varepsilon).
\end{align}
As in the case of the single delta function \eqref{boundary values}, the difference of the boundary values $F_+(x)-F_-(x)$ yields an array of simple poles with period $\pi$ on $\mathbb{R}$,
\begin{align}\label{regularized periodic delta}
    F_+(x)-F_-(x)&=\frac{i}{2\pi}(\cot(x+i\varepsilon)-\cot(x-i\varepsilon))=\frac{i}{2\pi}\sum_n\left(\frac{1}{x+n\pi+i\varepsilon}-\frac{1}{x+n\pi-i\varepsilon}\right)\nonumber\\
    &=-\frac{1}{2\pi}\left(\frac{e^{ix+\varepsilon}+e^{-ix-\varepsilon}}{e^{ix+\varepsilon}-e^{-ix-\varepsilon}}
    -\frac{e^{ix-\varepsilon}+e^{-ix+\varepsilon}}{e^{ix-\varepsilon}-e^{-ix+\varepsilon}}\right)\nonumber\\
    &=\frac{1}{\pi}\frac{\sinh2\varepsilon}{\cosh2\varepsilon-\cos2x}.
\end{align}
This final expression can be identified, within Sato's hyperfunction theory, with the regularised periodic delta function of period $\pi$.
Adjusting the period with $2\pi/\mu_0$, we obtain the formula \eqref{periodic delta}.
Note that the regularization parameter of $S^2_\varepsilon$ is identified with the $\varepsilon$ introduced here to avoid singularities in the complex $z$-plane.
Although the integration appearing in the second Chern number $Q$ is performed over real variables, the introduction of the complex $x_0$-plane is necessary to reveal the delta function structure of the integrand.

\bibliographystyle{unsrt}

\bibliography{bibitem_cal2}

\end{document}